\documentclass[aps,prc,twocolumn,floatfix,showpacs,amsmath,amssymb]{revtex4}
\usepackage{graphicx}
\usepackage{dcolumn}
\usepackage{color} 
  \def\nuc#1#2{\relax\ifmmode{}^{#1}{\protect\text{#2}}\else${}^{#1}$#2\fi}
  \def\itnuc#1#2{\setbox\@tempboxa=\hbox{\scriptsize\it #1}
    \def\@tempa{{}^{\box\@tempboxa}\!\protect\text{\it #2}}\relax
    \ifmmode \@tempa \else $\@tempa$\fi}
  
  \newcommand{\beq}{\begin{equation}}
  \newcommand{\eeq}{\end{equation}}
  \newcommand{\bea}{\begin{eqnarray}}
  \newcommand{\eea}{\end{eqnarray}}

  \newcommand{\co}{(Color online)}

  \newcommand{\lisi}{\nuc{6}{Li}}
  \newcommand{\lise}{\nuc{7}{Li}}
  \newcommand{\bese}{\nuc{7}{Be}}
  \newcommand{\liei}{\nuc{8}{Li}}
  \newcommand{\lini}{\nuc{9}{Li}}
  \newcommand{\beni}{\nuc{9}{Be}}
  \newcommand{\bete}{\nuc{10}{Be}}
  \newcommand{\liel}{\nuc{11}{Li}}
  \newcommand{\beel}{\nuc{11}{Be}}
  \newcommand{\nm}{\ensuremath{N_\mathrm{max}}}
  \newcommand{\ho}{\ensuremath{\hbar \Omega}}
  \newcommand{\hoopt}{\ensuremath{\hbar \Omega_A}}
  \newcommand{\nn}{\ensuremath{N\!N}}
  
 \newcommand{\cdb}{CDB2k}
 
 \newcommand{\inoy}{INOY}
\begin{document}

\title{Charge radii and electromagnetic moments of {L}i and {B}e
  isotopes from the \emph{ab initio} no-core shell model}
\author{C. Forss\'en} \email[]{christian.forssen@chalmers.se}
\affiliation{Fundamental Physics, Chalmers University of Technology, 412
  96 G\"oteborg, Sweden} \author{E. Caurier} \affiliation{Institut de
  Recherches Subatomiques
  (IN2P3-CNRS-Universit\'e Louis Pasteur)\\
  Batiment 27/1, 67037 Strasbourg Cedex 2, France}
\author{P. Navr\'atil} \affiliation{Lawrence Livermore National
  Laboratory, P.O. Box 808, L-414, Livermore, CA 94551, USA}
\date{\today}
\begin{abstract}
  Recently, charge radii and ground-state electromagnetic moments of Li
  and Be isotopes were measured precisely.  We have performed
  large-scale {\it ab initio} no-core shell model calculations for these
  isotopes using high-precision nucleon-nucleon potentials.  The
  isotopic trends of our computed charge radii, quadrupole and
  magnetic-dipole moments are in good agreement with experimental
  results with the exception of the \liel\ charge radius. The magnetic
  moments are in particular well described whereas the absolute
  magnitudes of the quadrupole moments are about $10 \%$ too small. The
  small magnitude of the \lisi\ quadrupole moment is reproduced, and
  with the CD-Bonn \nn\ potential also its correct sign.
\end{abstract}
\maketitle
%
%
Recent developments of both experimental and theoretical techniques have
allowed for very precise measurements of charge radii and ground-state
electromagnetic moments of exotic
isotopes~\cite{san06:96,nor08:0809.2607,bor05:72,neu08:101}. In
particular, charge radii can be determined through measurements of
isotope shifts in certain atomic transitions using high-precision laser
spectroscopy at isotope separation
facilities~\cite{san06:96,nor08:0809.2607}. Theoretical calculations of
electron correlations, as well as relativistic and QED corrections, can
be performed~\cite{yan03:91,puc08:78} to yield very small
uncertainties in the extracted results. Electric quadrupole and magnetic
dipole moments can be determined using an experimental method that is
based on the nuclear magnetic resonance
technique~\cite{bor05:72,neu08:101}.

These observables reflect, in different ways, the evolving nuclear
structure along the isotopic chains. Fascinating trends with varying
$N/Z$ ratio have been revealed~\cite{jon04:389}. It is a true challenge
for theoretical methods to compute these observables, and to reproduce
all of the observed trends simultaneously. In particular, for \emph{ab
  initio} many-body methods this type of study allows to test properties
of the high-precision nuclear Hamiltonians that are used as the single
input to the calculations. But these calculations also constitute a
critical test of the limitations of the many-body method itself and of
the operators that are used to compute the matrix elements.

Efforts to compute some of these observables with microscopic approaches
have been performed using different realistic and semi-realistic
interactions~\cite{pie02:66,nav98:57,var02:66,kan01:142}. Concerning the
\emph{ab initio} no-core shell model (NCSM) used in this study,
increased computational capabilities and improved algorithms have led to
the opportunity to reach significantly extended model spaces. In
addition, studies of realistic nuclear Hamiltonians have led some
authors to explore the extent to which effects of multi-nucleon forces
can be absorbed by non-local terms in the \nn\
interaction~\cite{dol04:69,dol03:67}. The use of these \nn\ interactions
allows, to some extent, to study three-body interaction effects while
still maximizing the size of the model space.

In concert, these developments can be regarded as a strong motivation to
repeat and extend some earlier NCSM
studies~\cite{nav98:57,nav06:73,for05:71}. In this Rapid Communication
we calculate the charge radii and electromagnetic moments of the $A\leq
11$ chains of Li and Be isotopes. We compare the performance of two very
different \nn\ interactions: (i) the CD-Bonn 2000 interaction
(\cdb)~\cite{mac01:63}, that is a charge-dependent \nn\ interaction
based on one-boson exchange; and (2) the INOY IS-M~\cite{dol04:69} that
is a phenomenological interaction for which non-locality was introduced
in certain partial waves so that the binding energies of \nuc{3}{H} and
\nuc{3}{He} are described correctly~\cite{note09:inoy}. All calculations
are performed up to very large model spaces and efforts are made to
quantify the rates of convergence of the observables.
%
%

The NCSM method has been described in great detail in
several papers, see e.g., Refs.~\cite{nav00:84,nav00:62,nav01:87}. Here,
we just outline the approach as it is applied in the present study.
%
%
We start from the intrinsic two-body Hamiltonian for the $A$-nucleon
system $H_A = \mathcal{T}_\mathrm{rel} + \mathcal{V}$, where
$\mathcal{T}_\mathrm{rel}$ is the relative kinetic energy and
$\mathcal{V}$ is the sum of two-body nuclear and Coulomb
interactions. We solve the many-body problem in a large but finite
harmonic-oscillator (HO) basis truncated by a chosen maximal total HO
energy of the $A$-nucleon system.  If we use realistic nuclear
interactions such as \cdb\ or \inoy, that generate strong short-range
correlations, it is necessary to compute a model-space dependent
effective Hamiltonian. For this purpose, we perform a unitary
transformation~\cite{suz80:64,suz94:92} of the Hamiltonian, which
accommodates the short-range correlations. In general, the transformed
Hamiltonian is an $A$-body operator. Our simplest, yet nontrivial,
approximation is to develop a two-particle cluster effective
Hamiltonian, while the next improvement is to include three-particle
clusters, and so on. The effective interaction is then obtained from the
decoupling condition between the model space and the excluded space for
the two-nucleon transformed Hamiltonian.  Details of the procedure are
described in Refs.~\cite{nav00:84,nav00:62,nav00:61}.

\begin{figure}[hbt]
  \includegraphics*[width=0.95\columnwidth]
      {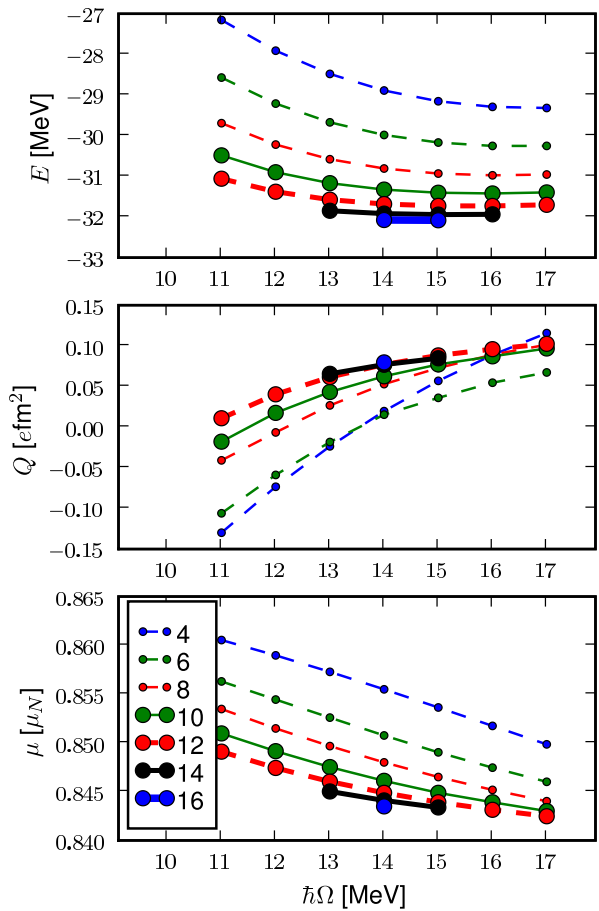}
\caption{\co\ $\hbar\Omega$-dependence for \lisi\ ground-state
  observables calculated with the \inoy\ interaction. Each
  curve corresponds to a particular model space represented by the
  parameter \nm\ (see text for details).%
  \label{fig:z3-hw}}
\end{figure}
In the present application we aim to describe nuclear observables that,
in some cases, are sensitive to long-range properties of the wave
function. It is therefore essential to include as many terms as possible
in the expansion of the total wave function. For this reason, we limit
ourselves to the use of realistic two-nucleon ($\nn$) interactions and
we only present results obtained with the two-body cluster
approximation.

Our calculations
are not variational, so higher order terms may contribute with either
sign to total binding. Hence, evaluating the dependence on the
basis-space parameters help calibrate our convergence.  Once the
effective interaction is derived, we diagonalize the effective
Hamiltonian in a Slater determinant HO basis that spans a complete
\nm\ho\ space. This is a highly non-trivial problem because of the very
large dimensions we encounter.  To solve this problem, we have used a
specialized version of the shell model code Antoine~\cite{cau99:30},
adapted to the NCSM~\cite{cau01:64}. 
%
The first step in our study is a search for the optimal HO
frequencies. The converged results should not depend on the HO frequency.
In practice however, due to the cluster approximation of the effective
interaction, our results will be sensitive to the choice of
\ho. Furthermore, by construction the effective interactions depend on
the size of the model space, \nm, and on the number of nucleons, $A$. In
order to investigate these dependences we have performed calculations
for a large series of frequencies. As an example, the results from this
study for the \lisi\ ground-state binding energy and electromagnetic
moments are presented in Fig.~\ref{fig:z3-hw} for the \inoy\
interaction.

We are looking for the region in which the dependence on \ho\ is the
smallest; and we select this frequency (from the calculation in the
largest model space) as a starting point for our detailed investigation
of ground-state observables. The selected optimal HO frequencies,
\hoopt, are presented in Table~\ref{tab:optfreq}.
\begin{table}[tbh]
  \caption{Selected optimal HO frequencies, \hoopt, (in [MeV]). Also
    shown is the maximum model space (represented by \nm)
    that was reached for each mass number $A$.
    \vspace*{1ex}%
    \label{tab:optfreq}}
  \begin{ruledtabular}
    \begin{tabular}{c|cccccc}
                 &  \multicolumn{6}{c}{Mass number} \\   
     Interaction & $6$ &  $7$ & $8$ & $9$ & $10$ & $11$    \\
      \hline
      \cdb\      & 11    & 11     & 12    & 12    & 13     
      & 13\footnote{The choice \hoopt=12~MeV is used for \liel.}  \\
      \inoy\     & 14    & 16     & 16    & 16    & 17     & 17  \\
      \hline \\[-0.5ex]
      \nm\       & 16    & 12     & 12    & 10    & 10     & 10
    \end{tabular}
  \end{ruledtabular}
\end{table}

In addition, our results will depend on the size of the model space
represented by the parameter \nm. Our approach is to fix the HO
frequency, at the value \hoopt\ specified in Table~\ref{tab:optfreq},
and observe the convergence with increasing model space. At the same
time, we investigate the \nm-dependence at $\hoopt \pm 1$~MeV.  

Using these studies we can gauge the degree of convergence of our final
results. For a given observable $O$ the total error $\Delta O$ is
estimated by
\beq
\Delta O = \sqrt{\Delta O_{N}^2 + \Delta O_\Omega^2},
\eeq
where $\Delta O_{N}$ and $\Delta O_\Omega$ represent the observed rates
of convergence with respect to model-space size and HO frequency,
respectively. We choose $\Delta O_{N}$ simply as the difference between
the numerical results in the largest and next-to-largest model spaces
using the optimal HO frequency. To determine $\Delta O_\Omega$ we
consider the triplet of results computed at $\hoopt$ and $\hoopt \pm
1$~MeV in the largest model space for which all three are available. We
then define the quantity $\Delta O_\Omega$ as the difference between the
maximum and minimum value of this triplet of results. The total error
$\Delta O$ is presented in the tables of this paper. Ground-state binding
energies ($E$) for Li isotopes are summarized in Table~\ref{tab:e}.
\begin{table}[bth]
  \caption{Ground-state binding energies ($E$) for Li
    isotopes. The \inoy\ results are extrapolated energies (with
    convergence-rate estimates), while the actual computed
    results from the largest model space are given
    in square brackets. The experimental results are from
    \cite{til02:708,til04:745,bac08:100}.
    \vspace*{1ex}%
    \label{tab:e}}
  \begin{ruledtabular}
    \begin{tabular}{r|dddd}
                 &  \multicolumn{4}{c}{$E$ [MeV]} \\[1ex]  
                 & \multicolumn{1}{c}{\cdb} & \multicolumn{2}{c}{\inoy} 
                 & \multicolumn{1}{c}{Exp} \\
      \hline
      \lisi      & 29.07(41) & 32.33(19) & [32.07] & 31.99 \\
      \lise      & 35.56(23) & 39.62(40) & [38.89] & 39.24 \\
      \liei      & 35.82(22) & 41.27(51) & [39.94] & 41.28 \\
      \lini      & 37.88(82) & 45.86(74) & [42.30] & 45.34 \\
      \liel      & 37.72(45) & 42.50(95)\footnote{The exponential
        convergence rate is not fully reached.} & [40.44] & 45.72(1) \\
    \end{tabular}
  \end{ruledtabular}
\end{table}
We note that for the \inoy\ NN potential, the ground-state energy
convergence is very uniform with respect to the HO frequency with
systematic changes with $N_{\rm max}$. The convergence with increasing
$N_{\rm max}$ is quite evident in particular for $A=6-9$, see e.g. the
top panel of Fig.~\ref{fig:z3-hw}, and we are able to extrapolate
assuming an exponential dependence on $N_{\rm max}$ as $E(N_{\rm
  max})=E_\infty+a \;{\rm exp}(-bN_{\rm max})$.  The same type of
extrapolation was successfully performed for the He
isotopes~\cite{cau06:73}.  Therefore, extrapolated \inoy\ energies are
given in Table~\ref{tab:e} together with computed results from the
largest model space (in square brackets).

It also deserves to be mentioned that we can, in principle, have
different \ho-dependence for different observables. In particular since
we're using effective Hamiltonians to compute the eigenenergies and
eigenfunctions but are employing bare operators to evaluate other
observables. In practice, however, the choice of \hoopt\ is relatively
stable. The exception to this general rule is observed for those
observables that depend critically on the size of the system. In
particular, for large systems we sometimes observe a different
\ho-dependence for the electric quadrupole moment and for the radius. In
this case the convergence-rate estimate should be viewed as providing
the magnitude of variations around the maximum model space reached.
The effective-operator formalism for general one- and two-body
operators was developed and studied in Ref.~\cite{ste05:71}, finding weak
renormalization of long-range operators.

To complete the present survey we have collected results for odd-$A$ Be
isotopes from earlier papers. The \bese\ results are partly from
Ref.~\cite{nav06:73} while $^{9,11}$Be results are from
Ref.~\cite{for05:71}. However, additional calculations were performed in
the present study to provide convergence rate estimates. In
addition, new calculations have been performed for the \bete\ ground
state.
%
%

\begin{figure}[htb]
  \includegraphics*[width=0.95\columnwidth]
      {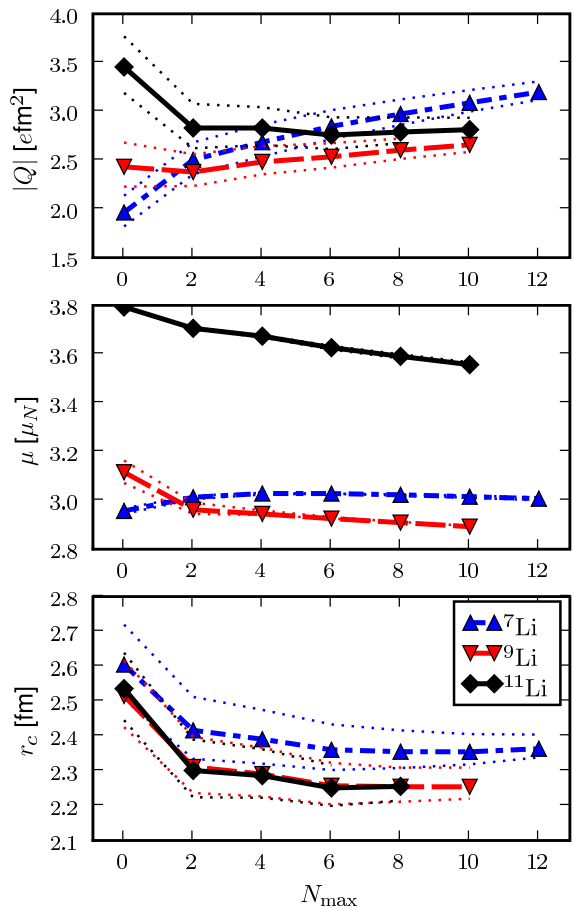}\\
      \caption{\co\ $N_{max}$-dependence of ground-state electromagnetic
        moments and charge radius for odd Li isotopes calculated with
        the \cdb\ interaction. Each curve corresponds
        to a fixed HO frequency. In particular, thick (thin) lines
        correspond to \hoopt\ ($\hoopt \pm 1$~MeV) for each isotope.%
  \label{fig:z3-nm-qmurc}}
\end{figure}

Ground-state quadrupole and magnetic dipole moments have been calculated
in the impulse approximation using bare operators. Two-body
meson-exchange currents have been shown to increase isovector magnetic
moments~\cite{mar08:78, sch98:58} but are not included here. The
\nm-dependence of $Q$ and $\mu$ is presented in
Fig.~\ref{fig:z3-nm-qmurc} for odd Li isotopes using the \cdb\
interaction. We note that the magnetic moment is usually very well
converged in the NCSM. The quadrupole moment is sensitive to the length
scale of the NCSM basis and for smaller model spaces it tends to
increase with decreasing HO frequency. 
\begingroup
\squeezetable
\begin{table}[tbh]
  \caption{Ground-state quadrupole moments ($Q$),
    magnetic dipole moments ($\mu$), and charge radii ($r_c$) for Li and Be
    isotopes. Theoretical results for \bese\ and \nuc{9,11}{Be} are
    partly from Refs.~\cite{nav06:73,for05:71}. The 
    experimental results are from
    Refs.~\cite{til02:708,til04:745,bor05:72,neu08:101} for 
    electromagnetic moments, and from Refs.~\cite{san06:96,
      nor08:0809.2607, puc08:78} for radii.
    \vspace*{1ex}%
    \label{tab:rcqmu}}
  \begin{ruledtabular}
    \begin{tabular}{r|ddd}
                 & \multicolumn{3}{c}{$Q$ [e fm$^2$]} \\[1ex]
                 & \multicolumn{1}{c}{\cdb} & \multicolumn{1}{c}{\inoy} 
                 & \multicolumn{1}{c}{Exp} \\
      \hline
      \lisi      & -0.066(40) & +0.080(19) & -0.0806(6) \\
      \lise      & -3.20(22)  & -2.79(17)  & -4.00(3)   \\
      \liei      & +2.78(12)  & +2.55(12)  & +3.14(2)   \\
      \lini      & -2.66(22)  & -2.30(12)  & -3.06(2)   \\
      \liel      & -2.81(27)  & -2.32(13)  & -3.33(5)   \\[1ex]
      \bese      & -5.50(48)  & -4.68(28)  & 
                    \multicolumn{1}{c}{--} \\
      \beni      & +4.12(26)  & +3.67(23)  & +5.288(38) \\[1ex]
      \hline
                 &  \multicolumn{3}{c}{$\mu$ [$\mu_N$]} \\[1ex]  
                 & \multicolumn{1}{c}{\cdb} & \multicolumn{1}{c}{\inoy} 
                 & \multicolumn{1}{c}{Exp} \\
      \hline
      \lisi      & +0.843(5) & +0.843(2)  & +0.822 \\
      \lise      & +3.01(2)  & +3.02(2)   & +3.256 \\
      \liei      & +1.24(6)  & +1.42(4)   & +1.654 \\
      \lini      & +2.89(2)  & +2.98(5)   & +3.437 \\
      \liel      & +3.56(4)  & +3.54(4)   & +3.671(1) \\[1ex]
      \bese      & -1.14(1)  & -1.15(1)   & -1.3995(5) \\
      \beni      & -1.22(9)  & -1.06(6)   & -1.1774 \\
      \beel      & -1.55(6)  & -1.47(3)   & -1.6813(5)
      \\[1ex]
      \hline
                 & \multicolumn{3}{c}{$r_c$ [fm]} \\[1ex]  
                 & \multicolumn{1}{c}{\cdb} & \multicolumn{1}{c}{\inoy} 
                 & \multicolumn{1}{c}{Exp} \\
      \hline
      \lisi      & 2.40(6)  & 2.29(4) & 2.540(28) \\
      \lise      & 2.36(7)  & 2.20(5) & 2.390(30)   \\
      \liei      & 2.31(8)  & 2.16(5) & 2.281(32) \\
      \lini      & 2.25(10) & 2.07(6) & 2.185(33) \\
      \liel      & 2.26(13) & 2.06(6) & 2.426(34) \\[1ex]
      \bese      & 2.56(10) & 2.39(5) & 2.647(17) \\
      \beni      & 2.41(11) & 2.22(6) & 2.519(12) \\
      \bete      & 2.34(9)  & 2.16(6) & 2.357(18) \\
      \beel      & 2.37(11) & 2.19(7) & 2.463(16) 
    \end{tabular}
  \end{ruledtabular}
\end{table}
\endgroup
Final results for all isotopes are presented in Table~\ref{tab:rcqmu}
together with recent and very precise experimental results using the
nuclear magnetic resonance technique~\cite{bor05:72,neu08:101}.  We note
that the two different interactions used in this study give very similar
isotopic trends but with a consistently smaller magnitude for the \inoy\
interaction. This observation is connected to the anomalously large
nuclear density generated by this interaction found already in
\nuc{4}{He} calculations~\cite{cau06:73,laz04:70}. For \lisi\ the
quadrupole moment is basically converged and the magnitude agrees well
with the experimental value. However, for the larger systems the
calculated magnitudes are consistently $\gtrsim 10\%$ too small as
compared to the experimental results. This result is expected from
working in a finite HO basis and we observe that the magnitude of the
quadrupole moment is steadily increasing with increasing model space, as
shown in Fig.~\ref{fig:z3-nm-qmurc}.
%
%

The point-nucleon radii ($r_{pt-p}$ and $r_{pt-n}$) can be computed from
the NCSM wave functions with complete removal of spurious CM motion. We
have translated these results into a rms charge radius ($r_c$) by
folding in proton $\langle R_p^2 \rangle$ and neutron $\langle R_n^2
\rangle$ rms charge radii~\cite{ams08:667} and adding the Darwin-Foldy
term~\cite{fri97:56} according to the relation $ \langle r_c^2 \rangle =
\langle r_{pt-p}^2 \rangle + \langle R_p^2 \rangle + N \langle R_n^2
\rangle / Z + 3\hbar^2/(4 M_p^2 c^2)$.
%
%
These results are compared to recent high-precision laser spectroscopy
measurements~\cite{san06:96,nor08:0809.2607} in Table~\ref{tab:rcqmu}. The
experimental results are obtained using previously measured \lise\ and
\beni\ radii as references. The convergence of the calculated radius
with increasing \nm\ is shown in Fig.~\ref{fig:z3-nm-qmurc} for odd Li
isotopes using the \cdb\ interaction. We also conclude that the larger
binding energy obtained with the \inoy\ interaction can partially be
explained by an abnormally large nuclear density and consequently a
charge radius that is too small. For \cdb, the radius results are rather
well converged and show good agreement with experiment. However, the
known halo structure of the \liel\ ground-state is not reproduced in the
limited HO model space of the NCSM. 
%

In summary we have computed charge radii and electomagnetic moments of
Li and Be isotopes using two different, high-precision nuclear
Hamiltonians within the \emph{ab initio} NCSM.  
\begin{figure}[hbt]
  \includegraphics*[width=0.95\columnwidth]
      {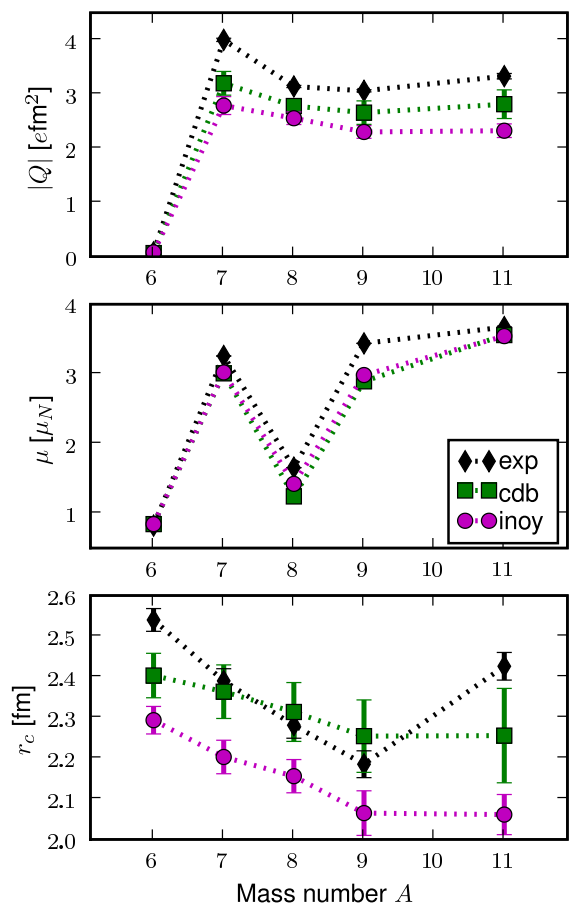}
\caption{\co\ NCSM calculated
  electric quadrupole moments, magnetic dipole moments, and charge radii
  of Li isotopes compared with experimental
  results. See also Table~\ref{tab:rcqmu}.%
  \label{fig:liA}}
\end{figure}
In Fig.~\ref{fig:liA} we compare the calculated and experimental trends
for these observables for the Li chain of isotopes. With the exception
of the radius of the \liel\ halo ground-state we find a very good
agreement between NCSM results and recent experiments. The overall
trends of all observables are well reproduced. Magnetic dipole moments
are usually characterized by very good convergence properties in the
NCSM and we find a good agreement with the experimental values. Another
success is the tiny quadrupole moment of \lisi\ that is known to pose a
difficult task for most theoretical calculations. In particular, the
general failure of three-body models for this observable has been blamed
on missing antisymmetrization of the valence nucleons and the nucleons
in the alpha-core~\cite{unk91:11}. The NCSM correctly reproduces the
very small value, but with \cdb\ and \inoy\ giving different
signs. Simultaneously, the trend for the much larger moments of $A=7-11$
is nicely reproduced. We note that the ratio $Q(\liel)/Q(\lini)$ is
found to be very close to unity, as confirmed recently by very precise
experimental data~\cite{neu08:101}. This finding is obtained without a
very accurate description of the dilute halo structure of \liel; a
structural feature that we find would require an extension of the HO
basis used in the standard NCSM. Still, the decrease of the charge
radius of $A=6-9$ isotopes is reproduced, although the \inoy\
interaction gives too high nuclear densities.

We conclude by observing that the recent achievement of performing very
precise measurements of ground-state properties of exotic isotopes 
proves to be a very valuable tool in the quest for understanding the
nuclear interaction, the forms of relevant operators, as well as the
evolving structure of the nuclear many-body system.

  This research was supported by the Swedish Research Council and the
  Knut and Alice Wallenberg Foundation. 
  Prepared by LLNL under Contract DE-AC52-07NA27344.
  This work was supported by the LDRD contract No.~PLS-09-ERD-020, by the
  U.S.\ DOE/SC/NP (Work Proposal Number SCW0498) and by the UNEDF SciDAC
  Collaboration under DOE grant DE-FC02-07ER41457.
  One of us (C.F.)
  acknowledges financial support from Stiftelsen Lars Hiertas Minne and
  from Stiftelsen L\"angmanska Kulturfonden.
%


\begin{thebibliography}{36}
\bibitem[{not()}]{note09:inoy}
\bibinfo{note}{It is important to note that, for this particular version, the
  on-shell properties of the triplet $P$-wave interactions were modified in
  order to improve the description of $3N$ analyzing powers.}
\end{thebibliography}
\end{document}